\documentclass{article}
\usepackage{spconf,amsmath,graphicx,bm,amssymb}

\usepackage{algorithmicx}
\usepackage[ruled]{algorithm}
\usepackage{algpseudocode}
\usepackage{algpascal}

\usepackage[colorlinks,
            linkcolor=red,
            anchorcolor=blue,
            citecolor=green]{hyperref}
\graphicspath{{figs/}}

\usepackage{booktabs}
\usepackage{multirow}

\def\b{{\bm{b}}}
\def\d{{\bm{d}}}
\def\u{{\bm{u}}}
\def\f{{\bm{f}}}

\def\W{{\bm{W}}}
\def\Y{{\bm{Y}}}

\def\L{{\mathcal{L}}}
\def\P{{\bm{\mathcal{P}}}}

\usepackage{xkeyval,xcolor}
\makeatletter
\newlength{\sfp@hseplen}\newlength{\sfp@vseplen}
\define@cmdkey{subfigpos}[sfp@]{pos}[ul]{}
\define@cmdkey{subfigpos}[sfp@]{font}[\small]{}
\define@cmdkey{subfigpos}[sfp@]{vsep}[0.8\baselineskip]{\setlength{\sfp@vseplen}{\sfp@vsep}}
\define@cmdkey{subfigpos}[sfp@]{hsep}[2pt]{\setlength{\sfp@hseplen}{\sfp@hsep}}
\newcommand{\subfigimg}[3][,]{%
  \setkeys{Gin,subfigpos}{pos,font,vsep,hsep,#1}
  \setbox1=\hbox{\includegraphics{#3}}
  \ifnum\pdfstrcmp{\sfp@pos}{ul}=0
    \leavevmode\rlap{\usebox1}
    \rlap{\hspace*{\sfp@hsep}\raisebox{\dimexpr\ht1-\sfp@vsep}{\sfp@font{#2}}}
    \phantom{\usebox1}
  \else\ifnum\pdfstrcmp{\sfp@pos}{ur}=0
    \leavevmode\usebox1
    \llap{\raisebox{\dimexpr\ht1-\sfp@vsep}{\sfp@font{#2}}\hspace*{\sfp@hsep}}
  \else\ifnum\pdfstrcmp{\sfp@pos}{lr}=0
    \leavevmode\usebox1
    \llap{\raisebox{\sfp@vsep}{\sfp@font{#2}}\hspace*{\sfp@hsep}}
  \else
    \leavevmode\rlap{\usebox1}
    \rlap{\hspace*{\sfp@hseplen}\raisebox{\sfp@vsep}{\sfp@font{#2}}}
    \phantom{\usebox1}
  \fi\fi\fi
}
\makeatother
\title{JSR-Net: A Deep Network for Joint Spatial-Radon Domain CT Reconstruction from incomplete data}
\twoauthors
 {Haimiao Zhang 
 \sthanks{The work of this author is funded by China Postdoctoral Science Foundation under grant 2018M641056.}, 
 Bin Dong\sthanks{Corresponding author. The research of this author is supported by NSFC grant 11831002. Email: dongbin@math.pku.edu.cn }}
	{Peking University,
	Beijing, China}
 {Baodong Liu\sthanks{Corresponding author. This work is supported by the National Key Scientific Instrument and Equipment Development Project 2017YFF0107200.}}
	{Chinese Academy of Sciences, Beijing, China,\\ University of  Chinese Academy of Sciences, \\Beijing, China}
\begin{document}
\maketitle
\begin{abstract}
CT image reconstruction from incomplete data, such as sparse views and limited angle reconstruction, is an important and challenging problem in medical imaging. This work proposes a new deep convolutional neural network (CNN), called JSR-Net, that jointly reconstructs CT images and their associated Radon domain projections. JSR-Net combines the traditional model-based approach with deep architecture design of deep learning. A hybrid loss function is adapted to improve the performance of the JSR-Net making it more effective in protecting important image structures. Numerical experiments demonstrate that JSR-Net outperforms some latest model-based reconstruction methods, as well as a recently proposed deep model.
\end{abstract}
\begin{keywords}
Joint spatial-Radon domain reconstruction, Sparse-view CT, Limited angle CT, Convolutional neural networks, Deep learning
\end{keywords}
\section{Introduction}
\label{sec:intro}

Deep learning is widely used in Natural language processing, speech recognition, computer vision and many other fields in recent years \cite{lecun2015deep}. Deep models have surpassed traditional handcrafted models and human experts in many tasks in imaging science such as image classification \cite{Krizhevsky2012ImageNet,he2015delving}. Recently, deep learning has attracted much attention from the medical imaging community \cite{wang2018image}.

In this work, we focus on the problem of CT image reconstruction from incomplete data \cite{natterer2001mathematics}, i.e., sparse views and limited angle reconstruction problems. Traditional CT image reconstruction algorithms include filtered back-projection(FBP) \cite{natterer2001mathematics}, algebraic reconstruction technique(ART) \cite{andersen1989algebraic} and model-based iterative reconstruction\cite{Jia2011GPU,chen2013limited}. However, these traditional methods cannot effectively utilize large image data sets, which limits their performance in various image reconstruction tasks.

The CNN based models are capable of learning multi-scale image features from large data sets with a cascade of simple modules. Although training a deep CNN can be time-consuming, it is very efficient to use in the validation stage. In \cite{yang2016deep}, the authors parameterized the proximal operator in the alternating direction method of multipliers (ADMM) algorithm using CNN for magnetic resonance imaging (MRI). Their deep model benefits from both the handcrafted image reconstruction model and CNN and is superior to some of the regularization based MR image reconstruction models. In \cite{adler2018learned}, the authors unrolled the primal-dual hybrid gradient(PDHG) algorithm to form a feed-forward deep network for CT image reconstruction. Then, the proximal operator and hyper-parameters in the PDHG algorithm were both approximated by CNN learned from the training data set. This new approach provides significant improvements in imaging quality when compared to the TV based variational model. A similar idea was also adopted to learn an iterative scheme for the nonlinear inverse problem of CT imaging \cite{Adler2017Solving}.

To suppress the artifacts induced by the incomplete data and noise, \cite{Dong2013} proposed a joint spatial-Radon domain reconstruction(JSR) model for sparse view CT imaging. This model can better guarantee the data consistency in the Radon domain which leads to better image quality. Similar idea to the JSR model is used later for positron emission tomography \cite{burger2014total}. To further protect image features, \cite{zhan2016ct} adopted a data-driven regularization in the JSR model. More recently, a re-weighting strategy was introduced to the JSR model to reduce the metal artifacts in multi-chromatic energy CT \cite{zhang2018reweighted}.

In view of the advantages of CNN for unrolled iterative scheme learning and the JSR model in CT image reconstruction, we propose a new deep architecture combining deep CNN with JSR model. This newly proposed model will be referred to as the JSR-Net. In order to fully exploit the advantage of the JSR-Net, a hybrid loss function is introduced to protect important image structures in the reconstructed images further. The JSR-Net is then validated and compared on large CT image data set with simulated projection data for both sparse view and limited angle problem.

\section{ Method }\label{sec:jsr_model_and_jsrnet}
In this section, we first review the JSR model and its associated iterative optimization algorithm. Then, we present the proposed JSR-Net that is inspired by the iterative optimization algorithm.

\subsection{The JSR Model and Algorithm}

The JSR model proposed in \cite{Dong2013} reads as follows
\begin{small}
\begin{eqnarray}\label{JSR-classical}
\min_{\bm{u},\f} \mathcal{F}(\u,\f,\Y)+
\|\bm{\lambda}_{1}\bm{W}_{1}\u\|_{1,2}+\|\bm{\lambda}_{2}\bm{W}_{2}\f\|_{1,2},
\end{eqnarray}
\end{small}
where the data fidelity term is defined by 
\begin{small}
$\mathcal{F}(\u,\f,\Y)=\frac{1}{2}\|R_{\bm{\Gamma}^{c}}(\f-\Y)\|^{2}+\frac{\alpha}{2}\|R_{\bm{\Gamma}}(\P\u-\f)\|^{2}+\frac{\gamma}{2}\|R_{\bm{\Gamma}^{c}}(\P\u-\Y)\|^{2}$
\end{small}, and $ R_{\bm{\Gamma}}$ is the restriction operator with respect to the domain with missing data $\bm{\Gamma}$. Here, $\bm{\Gamma}^{c}$ denotes the complement of $\bm{\Gamma}$ and it indicates the region with measured data.  $\P$ is the Radon transform, $\Y$ is the measured projection data.
The parameters $\alpha$ and $\gamma$ are properly chosen to balance the data consistency in Radon and spatial domain.
$\bm{W}_{i}, i=1,2,$ is the sparsity promoting transform such as tight wavelet frame transform \cite{Dong2010IASNotes}, and $\|\cdot\|$ is the $\ell_{2}$ norm.
$\bm{\lambda}_{i},i=1,2,$ is the multi-indexed regularization parameter to balance the sparsity prior and the data fidelity. The special norm $\|\cdot\|_{1,2}$ was firstly introduced in \cite{image2012cai}.

The optimization problem in \eqref{JSR-classical} can be solved by the ADMM algorithm with multi-blocks by introducing the augmented Lagrangian function :
\begin{footnotesize}
\begin{eqnarray}\label{JSR-augment}
\mathcal{L}(\bm{X},\bm{d},\bm{b})=\mathcal{F}(\bm{X},\Y)+\mathcal{R}(\bm{d})-\langle \bm{b}, \bm{W}\bm{X}-\bm{d}\rangle
+\frac{1}{2}\| \bm{WX}-\bm{d}\|_{\bm{\mu}}^{2}
\end{eqnarray}
\end{footnotesize}
where
\begin{footnotesize}
$\bm{d}=\left(\begin{matrix}
\bm{d}_{1}\\
\bm{d}_{2}
\end{matrix}
\right),$
\end{footnotesize}
\begin{footnotesize}
$\bm{X}=\left(\begin{matrix}
\bm{u}\\
\bm{f}
\end{matrix}
\right),$
\end{footnotesize}
\begin{footnotesize}
$\bm{\mu}=\left(\begin{matrix}
\mu_{1}\\
\mu_{2}
\end{matrix}
  \right)>\bm{0},$
\end{footnotesize}
and
\begin{footnotesize}
$\bm{W}=\left(\begin{matrix}
\bm{W}_{1}\\
\bm{W}_{2}
\end{matrix}
\right)$
\end{footnotesize}.
Here, 
\begin{footnotesize}
$\bm{b}=\left(\begin{matrix}
\bm{b}_{1}\\
\bm{b}_{2}
\end{matrix}
\right)$ 
\end{footnotesize}
is the dual variable.
Furthermore,
\begin{footnotesize}
$\mathcal{R}(\bm{d})=\|\bm{\lambda}_{1}\cdot\bm{d}_{1}\|_{1,2}+\|\bm{\lambda}_{2}\cdot\bm{d}_{2}\|_{1,2}$ 
\end{footnotesize}
and 
\begin{footnotesize}
$\| \bm{WX}-\bm{d}\|_{\bm{\mu}}^{2}=\mu_{1}\|\W_{1}\u-\d_{1}\|^{2}+\mu_{2}\|\W_{2}\f-\d_{2}\|^{2}$
\end{footnotesize}.
Then the ADMM scheme with primal and dual variables updated alternatively can be computed with closed-form formula for each subproblem. More details on the transition from (1) to (2) can refer to \cite{Dong2013}.
We omit the detailed derivation and present the algorithm in Algorithm \ref{alg:JSR-model}.

\alglanguage{pseudocode}
\begin{algorithm}[h!]
\caption{ JSR algorithm for \eqref{JSR-augment}}
\label{alg:JSR-model}
\begin{algorithmic}[1]
\State \small{Initialization:} $\b_{1}^{0}=\b_{2}^{0}=0$,
\While{stop criterion is not met }\\
update $\u$:
\begin{small}
\begin{eqnarray}\label{JSR-alg-u-update}
\u^{k+1}&=& \mathcal{A}^{-1}\left[\alpha \P^{\top}R_{\bm{\Gamma}} \f^{k}+\mathcal{B}+ \mu_{1}\W_{1}^{\top}(\d_{1}^{k}-\b_{1}^{k}) \right] \notag\\
\d_{1}^{k+1}&=& \mathcal{T}_{\bm{\lambda}_{1}/\mu_{1}}(\W_{1}\u^{k+1}+\b_{1}^{k}) \notag\\
\b_{1}^{k+1}&=& \b_{1}^{k}+(\W_{1}\u^{k+1}-\d_{1}^{k+1})
\end{eqnarray}
\end{small}
where 
\begin{small}
$\mathcal{A}= P^{\top}(\alpha R_{\bm{\Gamma}}+\gamma R_{\bm{\Gamma}^{c}}) \P +\mu_{1}$ 
\end{small}
and 
\begin{small}
$\mathcal{B}=\gamma \P^{\top}R_{\bm{\Gamma}^{c}}\Y$
\end{small}
\\
update $\f$:
\begin{small}
\begin{eqnarray}\label{JSR-alg-f-update}
\f^{k+1}&=& \mathcal{C}^{-1}\left[\alpha R_{\bm{\Gamma}} \P\u^{k+1}+\mathcal{D} +\mu_{2}\W_{2}^{\top}(\d_{2}^{k}-\b_{2}^{k}) \right] \notag\\
\d_{2}^{k+1}&=& \mathcal{T}_{\bm{\lambda}_{2}/\mu_{2}}(\W_{2}\f^{k+1}+\b_{2}^{k}) \notag\\
\b_{2}^{k+1}&=& \b_{2}^{k}+(\W_{2}f^{k+1}-\d_{2}^{k+1})
\end{eqnarray}
\end{small}
where 
\begin{small}
$\mathcal{C}=\alpha R_{\bm{\Gamma}}+ R_{\bm{\Gamma}^{c}}+\mu_{2}$
\end{small}
 and 
\begin{small}
$\mathcal{D}=R_{\bm{\Gamma}^{c}}\Y.$
\end{small}
\EndWhile
\State Output: $\u^{\ast}$
\end{algorithmic}
\end{algorithm}

\subsection{JSR-Net}
The idea of the design of JSR-Net is to unroll Algorithm \ref{alg:JSR-model} and approximate some of the operators using neural networks. In this way, we form a deep feed-forward network which is the proposed JSR-Net. To be more precise, we use CNNs to approximate the inverse operators $\mathcal{A}^{-1}$ and $\mathcal{C}^{-1}$, and the thresholding operator $\mathcal{T}$ in \eqref{JSR-alg-u-update} and \eqref{JSR-alg-f-update}. For simplicity, we denote these CNNs as $\mathcal{N}(\cdot;\Theta)$. Then, we obtain the JSR-Net as shown in Algorithm \ref{alg:JSRNet}, where all the parameters of the CNNs and $\alpha, \gamma, \mu_{1}, \mu_{2}$ are trainable parameters.

We adopt the same CNN architecture (with a different set of learnable parameters that varies for each iteration $k$) to approximate the matrix inversions in both of the $\u^{k}$- and $\f^{k}$-subproblem. The thresholding operator that appears in the $\d_{1}^{k}$- and $\d_{2}^{k}$-subproblem are approximated by a CNN with the same architecture but different from that of the $\u^{k}$- and $\f^{k}$-subproblem. To be more precise, the matrix inversion in the $\u^{k}$-subproblem (same for the $\f^k$-subproblem) is approximated by a CNN with a 3 level DenseNet architecture \cite{huang2017densely} followed by an LM-ResNet structure \cite{pmlr-v80-lu18beyond}. At each level of DenseNet, the input is processed by the composition of a Convolution(Conv) layer followed by PReLU activation \cite{he2015delving}.
While for the thresholding operators of the $\d_{1}^{k}$- and $\d_{2}^{k}$-subproblem, the CNN is composed of 3 convolution layers that comprises consecutive operations ``Conv $\to$ PReLU $\to$ Conv ''.
Note that, it is possible to approximate $\d_{1}^{k}$ and $\d_{2}^{k}$ by choosing a single convolution layer with ReLU activation.

\alglanguage{pseudocode}
\begin{algorithm}[h!]
\caption{ JSR-Net}
\label{alg:JSRNet}
\begin{algorithmic}[1]
\State \small{Initialization:} $\b_{1},\b_{2}, \u, \f, \W_{1}, \W_{2}, \mathcal{N}(\cdot)$
\For{k=0:N }\\
update $\u$:
\begin{small}
\begin{eqnarray}\label{JSRNet-u-update}
\u^{k+1}&=& \mathcal{N}_u(\left[ \P^{\top}R_{\bm{\Gamma}} \f^{k},\mathcal{B},\W_{1}^{\top}(\d_{1}^{k}-\b_{1}^{k}) \right];\Theta^{k}_u) \notag\\
\d_{1}^{k+1}&=& \mathcal{N}_{d_1}(\W_{1}\u^{k+1}+\b_{1}^{k};\Theta^k_{d_1}) \notag\\
\b_{1}^{k+1}&=& \b_{1}^{k}+(\W_{1}\u^{k+1}-\d_{1}^{k+1})
\end{eqnarray}
\end{small}
where 
\begin{small}
$\mathcal{B}=\gamma \P^{\top}R_{\bm{\Gamma}^{c}}\Y$
\end{small}
\\
update $\f$:
\begin{small}
\begin{eqnarray}\label{JSRNet-f-update}
\f^{k+1}&=& \mathcal{N}_f(\left[ R_{\bm{\Gamma}} \P\u^{k+1}, \mathcal{D}, \W_{2}^{\top}(\d_{2}^{k}-\b_{2}^{k}) \right];\Theta_f^k) \notag\\
\d_{2}^{k+1}&=& \mathcal{N}_{d_2}(\W_{2}\f^{k+1}+\b_{2}^{k};\Theta^k_{d_2}) \notag\\
\b_{2}^{k+1}&=& \b_{2}^{k}+(\W_{2}\f^{k+1}-\d_{2}^{k+1})
\end{eqnarray}
\end{small}
where 
\begin{small}
$\mathcal{D}=R_{\bm{\Gamma}^{c}}\Y$
\end{small}
\EndFor
\State Output: $\u^{\ast}$
\end{algorithmic}
\end{algorithm}

\section{Network training}
\subsection{Loss function}
To prevent over-smoothing, we adopt a hybrid loss function that blends SSIM loss, MSE and semantic loss to guide the training process.
Our structure-semantic-$\ell_2$ (SS2) hybrid loss function is defined as
\begin{small}
\begin{equation}\label{loss-func}
\mathcal{L}_{SS2}=\theta_{1} \mathcal{L}_{SSIM}+ \mathcal{L}_{MSE} +\theta_{3}\mathcal{L}_{sem},
\end{equation}
\end{small}
where 
\begin{small}
$\mathcal{L}_{SSIM}=\sum (1-SSIM(\u_{rec},\u_{truth}))$ 
\end{small}
is the error summation of SSIM over mini batch with respect to the reconstructed image $\u_{rec}$ and ground truth $\u_{truth}$. $\mathcal{L}_{MSE}=\mathcal{L}_{S}+\theta_{2}\mathcal{L}_{R}$ is the $\ell_{2}$ loss in spatial domain and Radon domain defined by $\mathcal{L}_{S}=\sum \|\u_{rec}-\u_{truth}\|^2$, and $\mathcal{L}_{R}=\sum \|R_{\bm{\Gamma}^{c}}\P(\u_{rec}-\u_{truth})\|^{2}$, respectively.
$\mathcal{L}_{sem}$ is the $\ell_{2}$ norm of the difference of the level sets (i.e. image contents) of $\u_{rec}$ and $\u_{truth}$ as $\mathcal{L}_{sem}=\|sem(\u_{rec})-sem(\u_{truth})\|^{2}$.

In the training stage, the loss function \eqref{loss-func} is minimized by the stochastic gradient descent algorithm with adaptive moment estimation(ADAM)\cite{kingma2015adam}.
The ADAM optimizer is adopted with a cosine annealing learning rate, at training step $t$, defined by $\eta=\frac{0.001}{2}(1+cos(\pi \frac{t}{t_{max}}))$; the second raw moment parameter $\beta=0.99$ and the gradient norm is clipped by $1.0$. The maximum training step $t_{max}$ is set to $10000$ for all the experiments. The training is conducted on Tensorflow 1.3.0 with a Titan Xp GPU with memory 10.75G.

\subsection{Training database}\label{sub:training_database}

The real clinical data ``the 2016 NIH-AAPM-Mayo Clinic Low Dose CT Grand Challenge'' \cite{mccollough2017low}
from Mayo Clinic is used for the training and validation. The data set contains abdomen CT images from five patients. 
It contains 1684 slices of images with $512\times 512$ pixels. We randomly select 50 slices for validation and the rest for training.

\section{Experiments}
The simulated CT imaging system has 1024 detectors with fan beam geometry. For the sparse view CT reconstruction, 90 available views uniformly distribute over $360^{\circ}$.
For the limited angle CT reconstruction, 150 views are collected which are measured by $1^{\circ}$ per scanning step.
The measured projection data, denoted by $\Y$, is contaminated by additive white Gaussian noise with variance $\sigma=0.05$ (for simplicity). 
A more general model with Poisson-Gaussian noise data \cite{zheng2018pwls} will be considered in the future.

We compare the reconstruction results of JSR-Net to FBP, PD-Net \cite{adler2018learned}, JSR model(Algorithm \ref{alg:JSR-model}).
To have a fair comparison, we choose $N=5$ in JSR-Net (Algorithm \ref{alg:JSRNet}) so that it has the similar number of trainable parameters as PD-Net.
The parameters in Algorithm \ref{alg:JSR-model} are empirically tuned for optimal performance. All the hyper-parameters in JSR-Net are fixed for the simulations of both sparse view and limited angle.

The source code of the PD-Net is publicly available. We only modify the forward imaging operator $\P$ by composing it with the data restriction operator $R_{\bm{\Gamma}^{c}}$ to obtain the final imaging system. All the other hyper-parameters are set as default. To demonstrate the advantage of the JSR-Net, we also train the PD-Net using the SS2 hybrid loss function \eqref{loss-func}.

\subsection{Sparse view CT image reconstruction}
\begin{figure}[t]
\centering
\begin{tabular}{@{}p{0.29\linewidth}@{}p{0.29\linewidth}@{}p{0.29\linewidth}@{}}
\subfigimg[width=\linewidth,pos=ul,font=\color{blue!90!red}]{(a)}{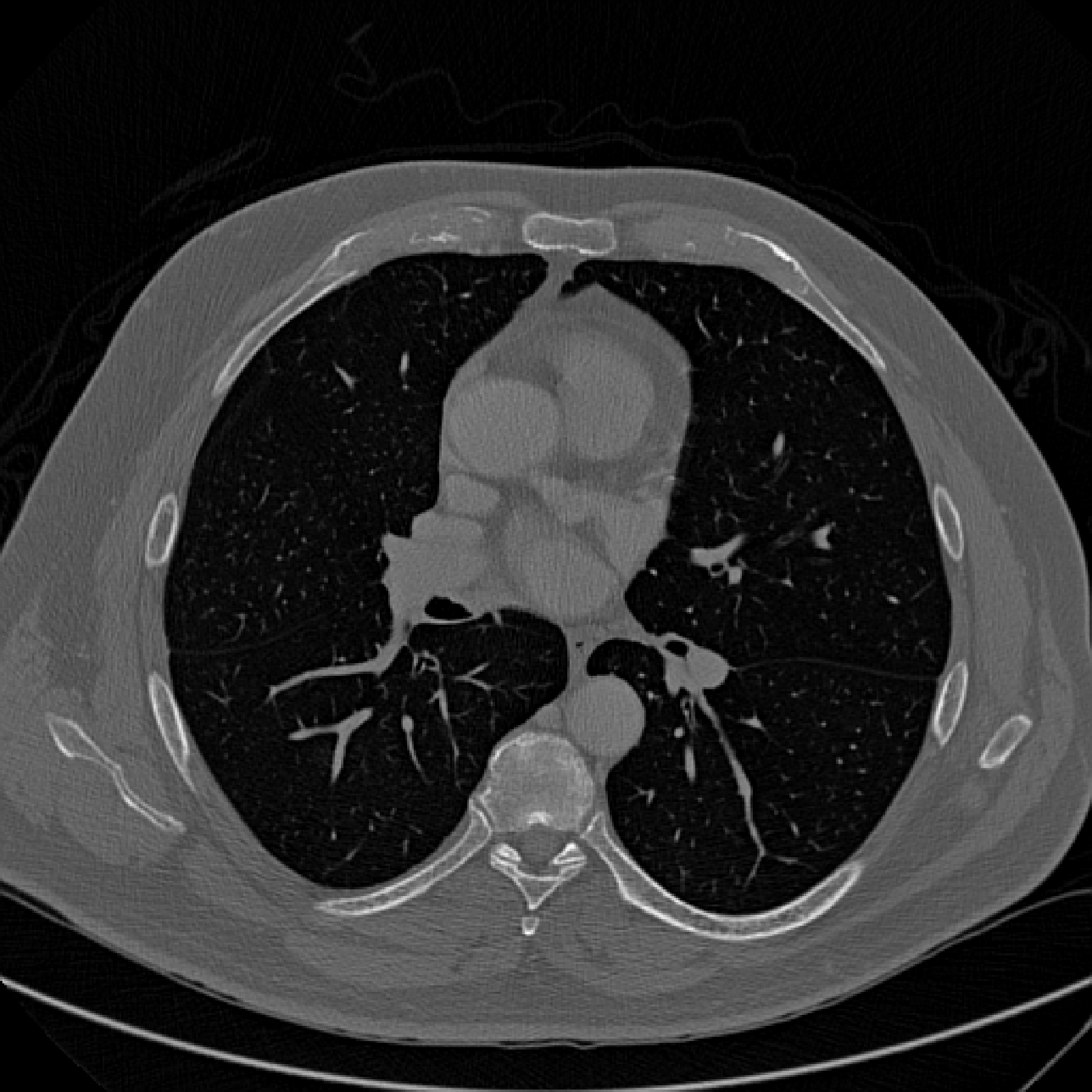} &
\subfigimg[width=\linewidth,pos=ul,font=\color{blue!90!red}]{(b)}{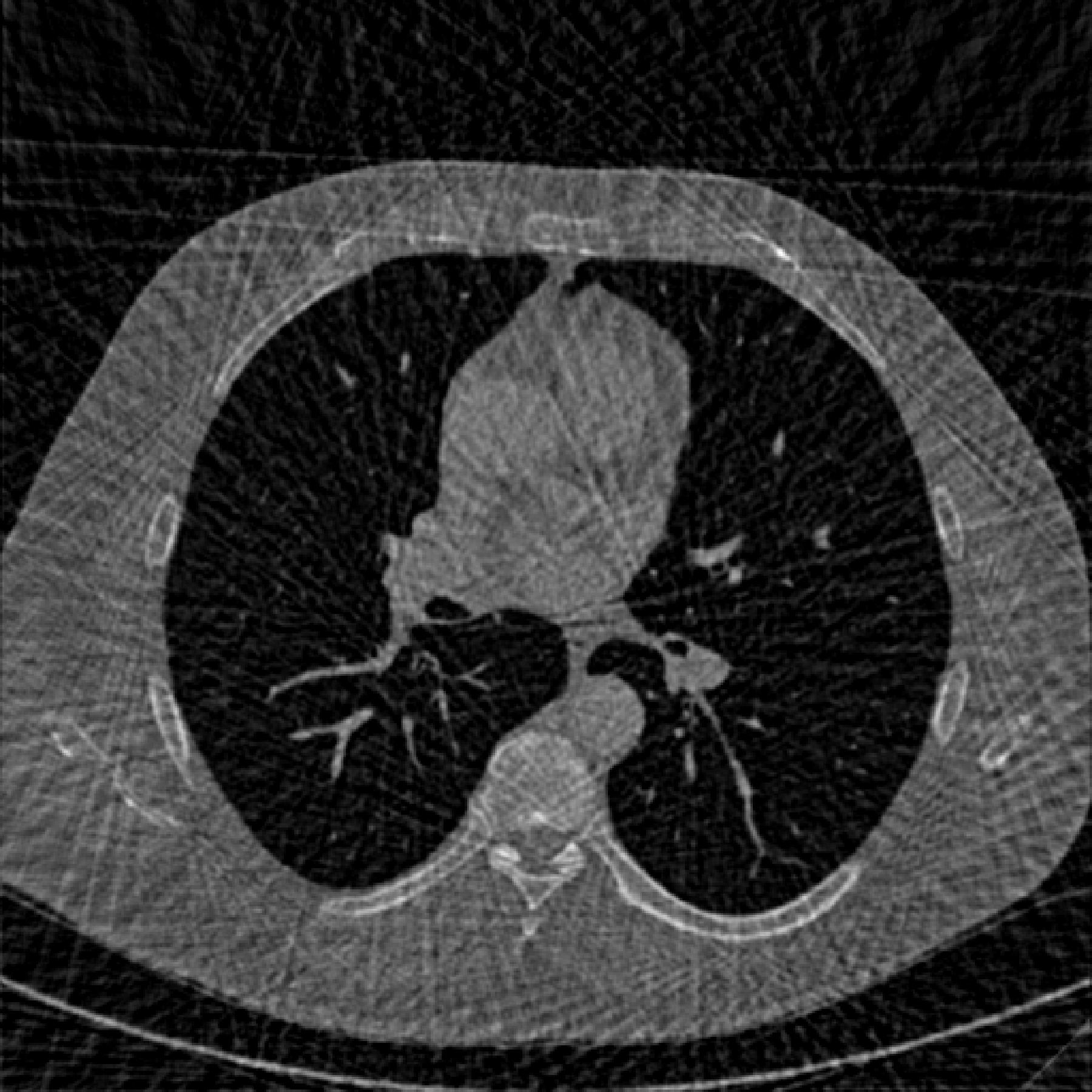}&
\subfigimg[width=\linewidth,pos=ul,font=\color{blue!90!red}]{(c)}{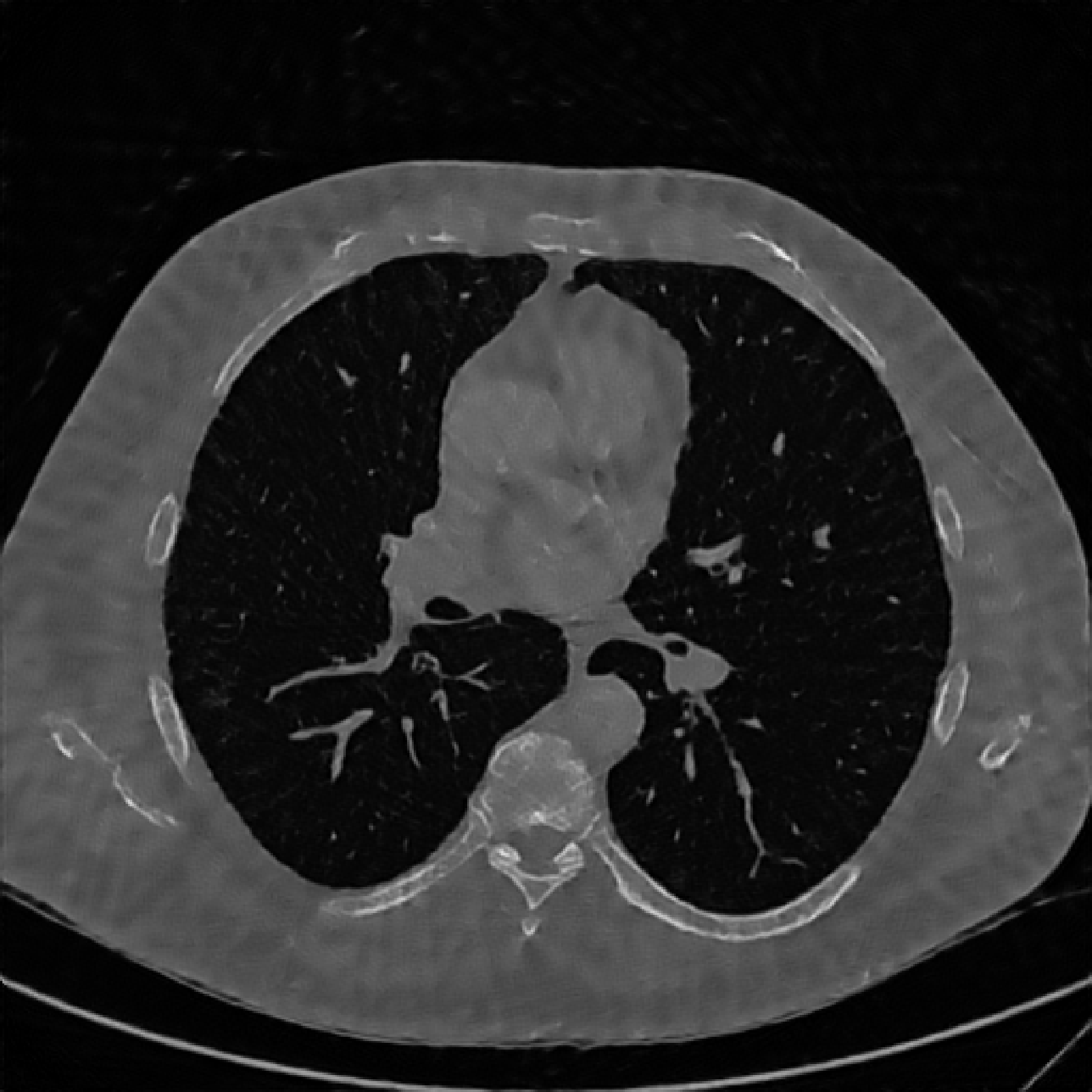}\\
\subfigimg[width=\linewidth,pos=ul,font=\color{blue!90!red}]{(d)}{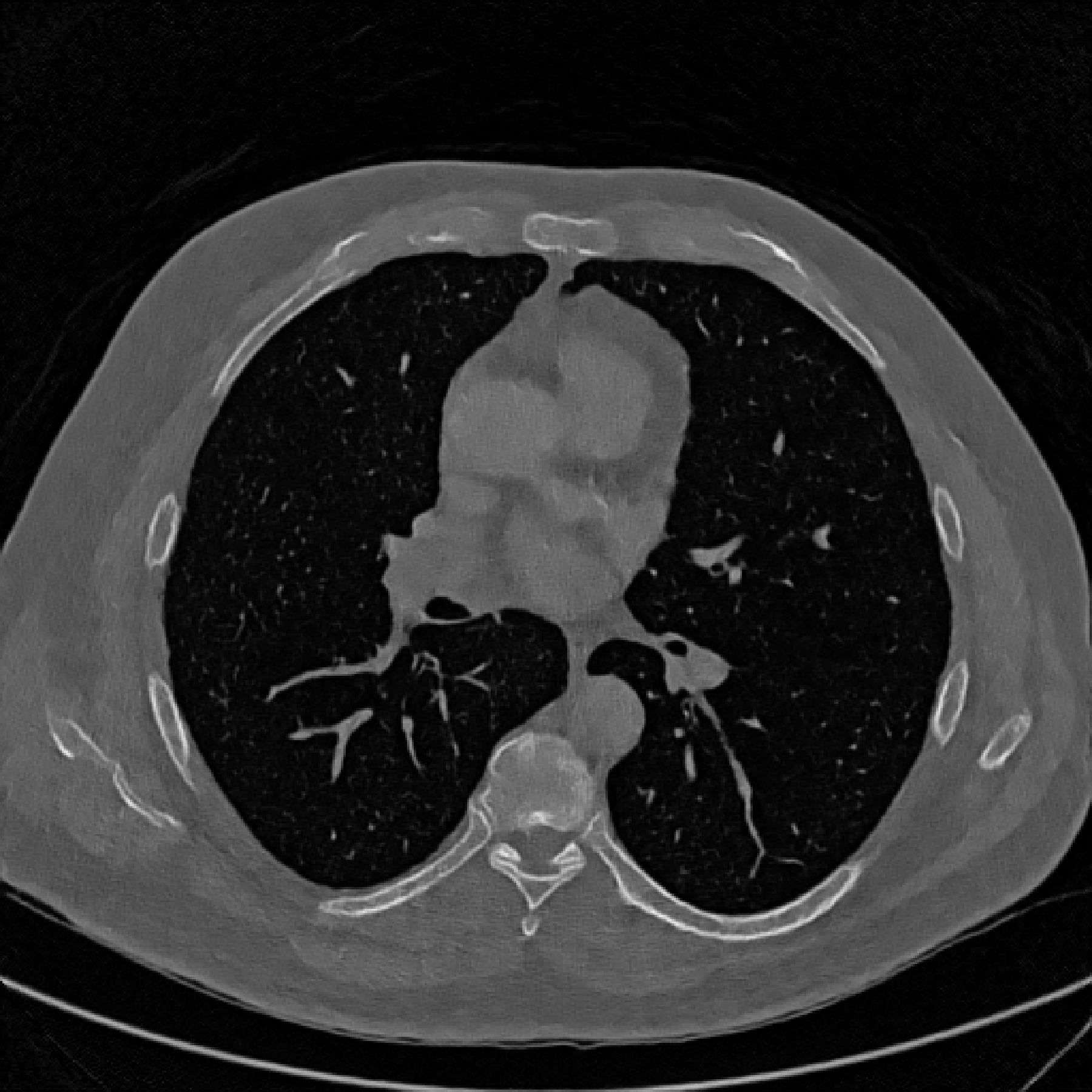}&
\subfigimg[width=\linewidth,pos=ul,font=\color{blue!90!red}]{(e)}{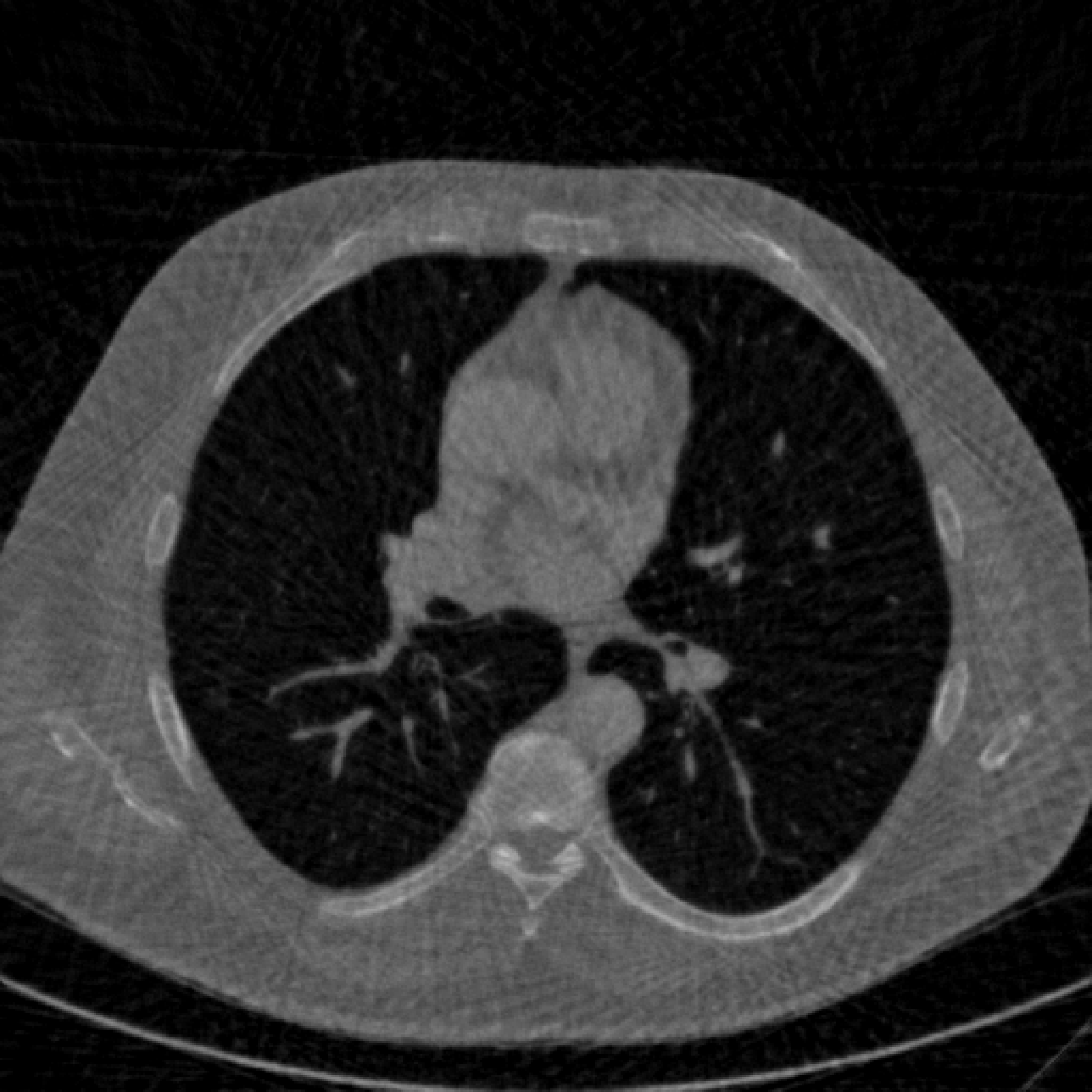}&
\subfigimg[width=\linewidth,pos=ul,font=\color{blue!90!red}]{(f)}{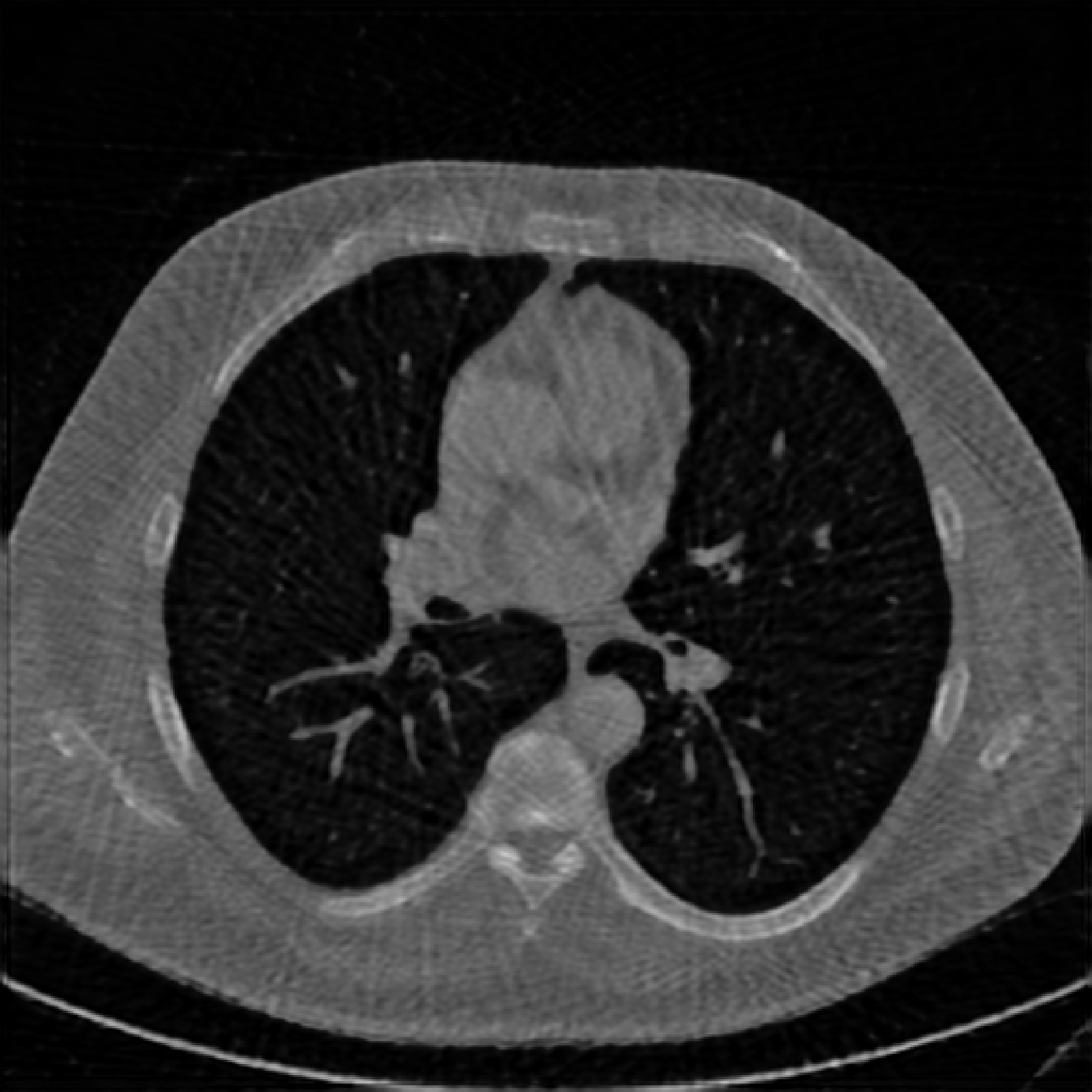}
 \\
\subfigimg[width=\linewidth,pos=ul,font=\color{blue!90!red}]{(g)}{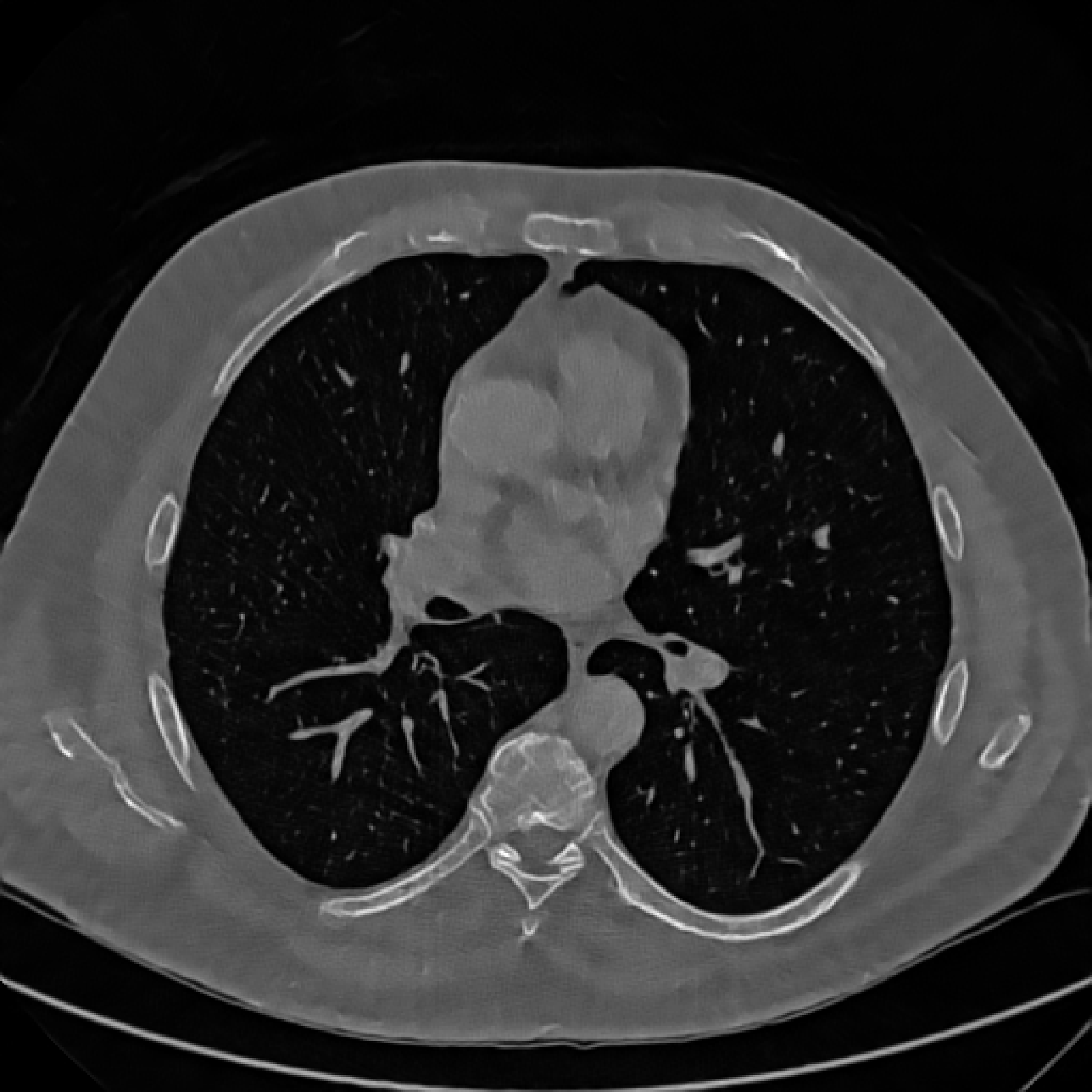}
 &
\subfigimg[width=\linewidth,pos=ul,font=\color{blue!90!red}]{(h)}{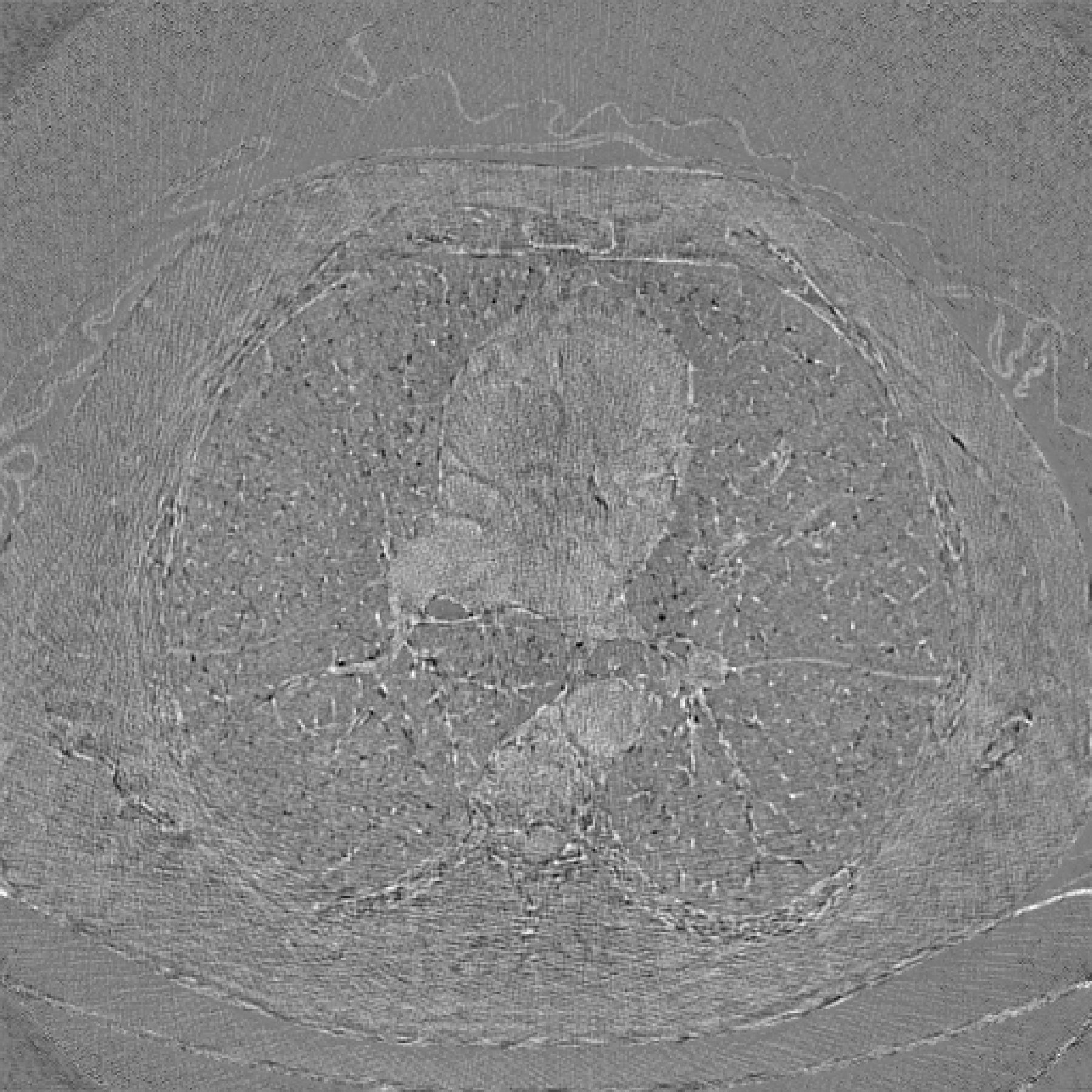}
&
\subfigimg[width=\linewidth,pos=ul,font=\color{blue!90!red}]{(i)}{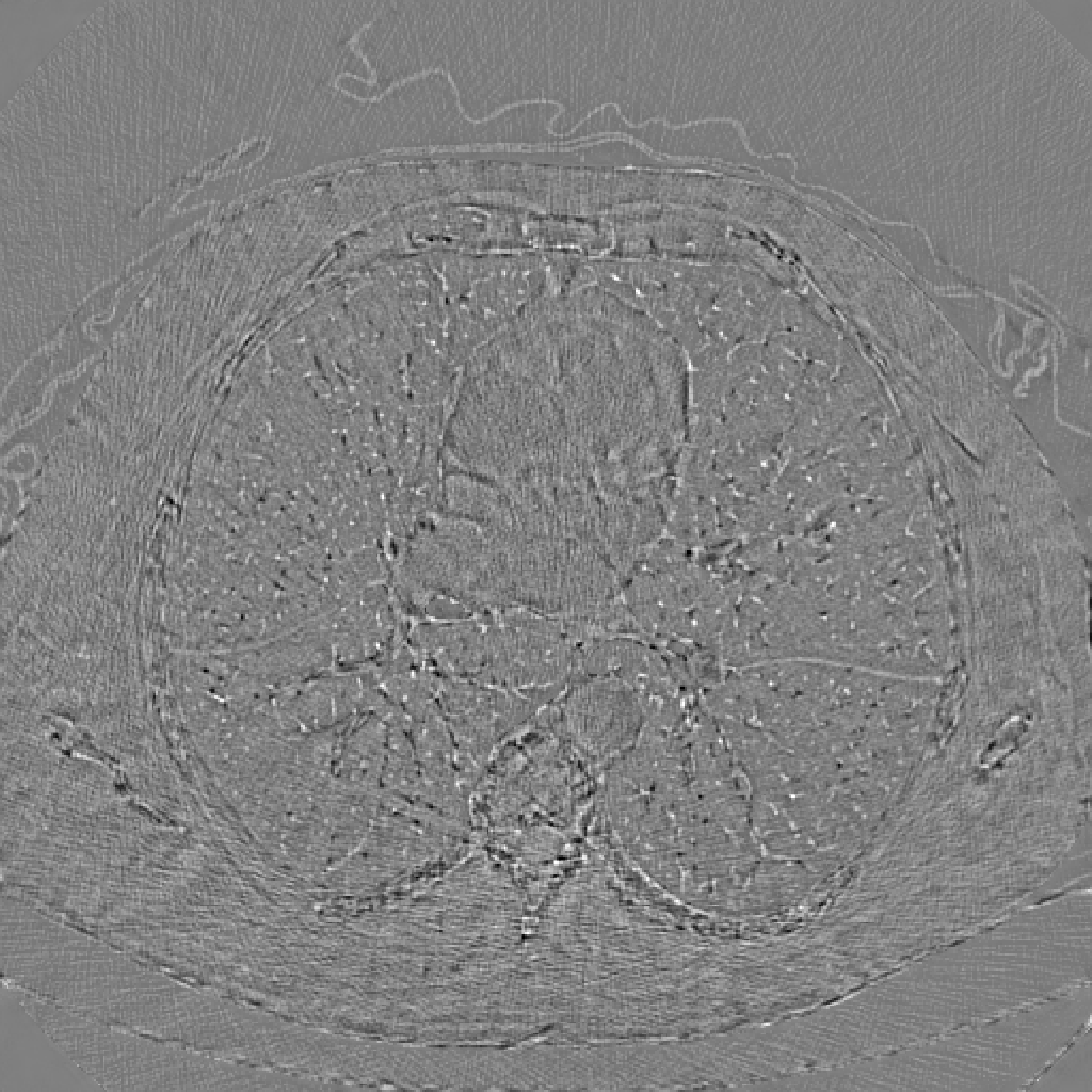}
\end{tabular}
\caption{Sparse view CT image reconstruction. (a)Ground truth; (b)FBP; (c)PD-Net, $\ell_{2}$; (d)PD-Net, SS2; (e)JSR model; (f) JSR-Net, $\ell_{2}$; (g)JSR-Net, SS2; (h)Error map of PD-Net, SS2; (i)Error map of JSR-Net, SS2.}
\label{fig:SACT}
\end{figure}

We evaluate the performance of the JSR-Net for sparse view CT image reconstruction. 
Fig. \ref{fig:SACT} shows an example of the ground truth and reconstructed images from FBP, PD-Net, JSR model(Algorithm \ref{alg:JSR-model}) and JSR-Net(Algorithm \ref{alg:JSRNet}). The gray scale window for FBP is set to [0, 0.5], and [0, 1] for the rest. The error maps are shown in Fig. \ref{fig:SACT}(h) and Fig. \ref{fig:SACT}(i).

The reconstruction results in Fig. \ref{fig:SACT} shows that both the JSR-Net and PD-Net outperform FBP and can better preserve sharp features than the JSR model.
The PD-Net with the proposed SS2 hybrid loss function \eqref{loss-func} has better performance than using merely the $\ell_{2}$ loss which was originally adopted by \cite{adler2018learned}. The JSR-Net with the SS2 hybrid loss function has the best overall performance, and it performs better than using only the $\ell_2$ loss in term of suppressing streak artifacts. Note that, since the semantic term $\L_{sem}$ is not crucial and empirically has little performance improvement for sparse view problem, we set $\theta_3=0$ in this experiment.
Error maps (with grayscale window [-0.2, 0.2]) in Fig. \ref{fig:SACT}(h) and (i) show that JSR-Net can better reconstruct image details than PD-Net.

\subsection{Limited angle CT image reconstruction}
\begin{figure}[h]
\centering
\begin{tabular}{@{}p{0.28\linewidth}@{}p{0.28\linewidth}@{}p{0.28\linewidth}@{}}
\subfigimg[width=\linewidth,pos=ul,font=\color{blue!90!red}]{(a)}{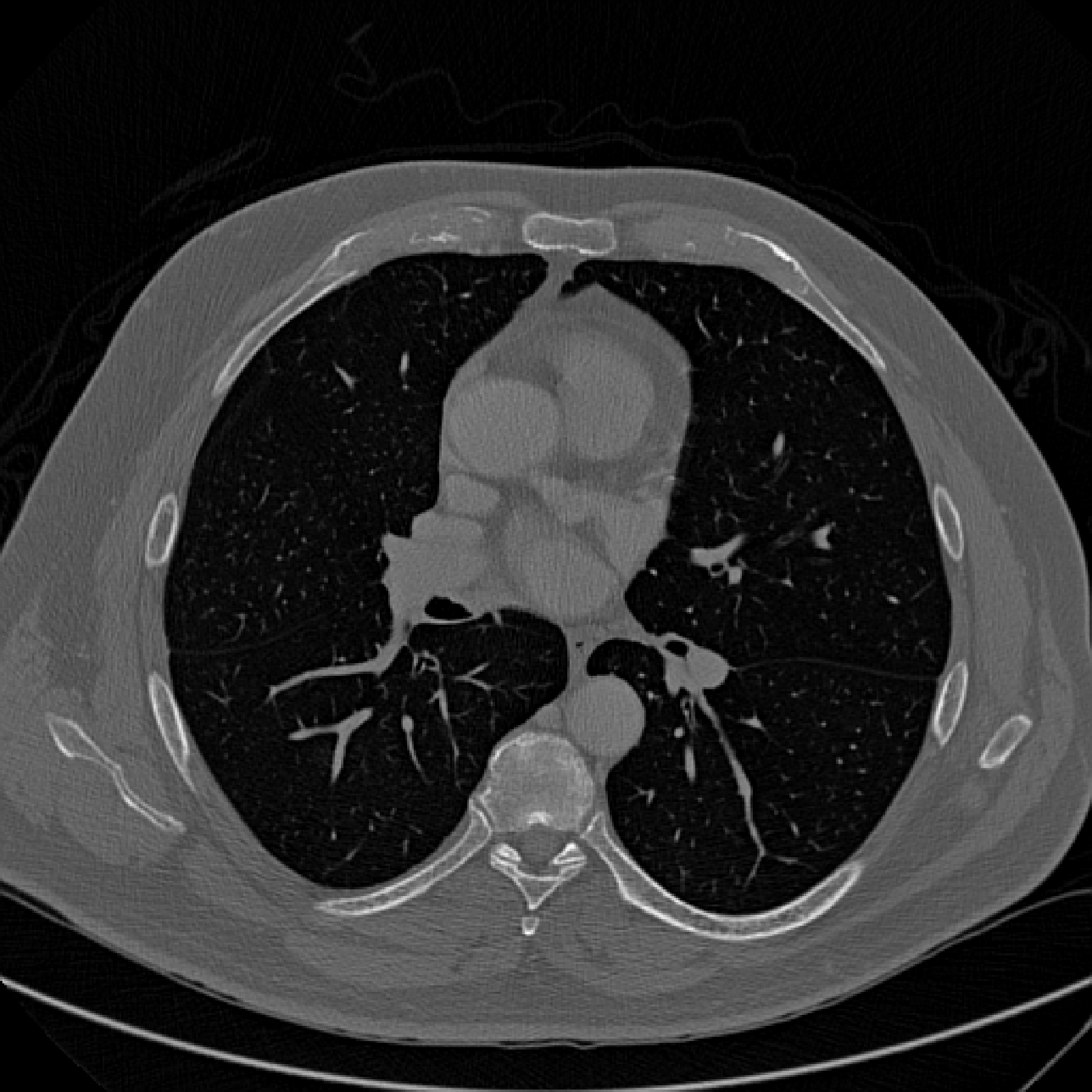} &
\subfigimg[width=\linewidth,pos=ul,font=\color{blue!90!red}]{(b)}{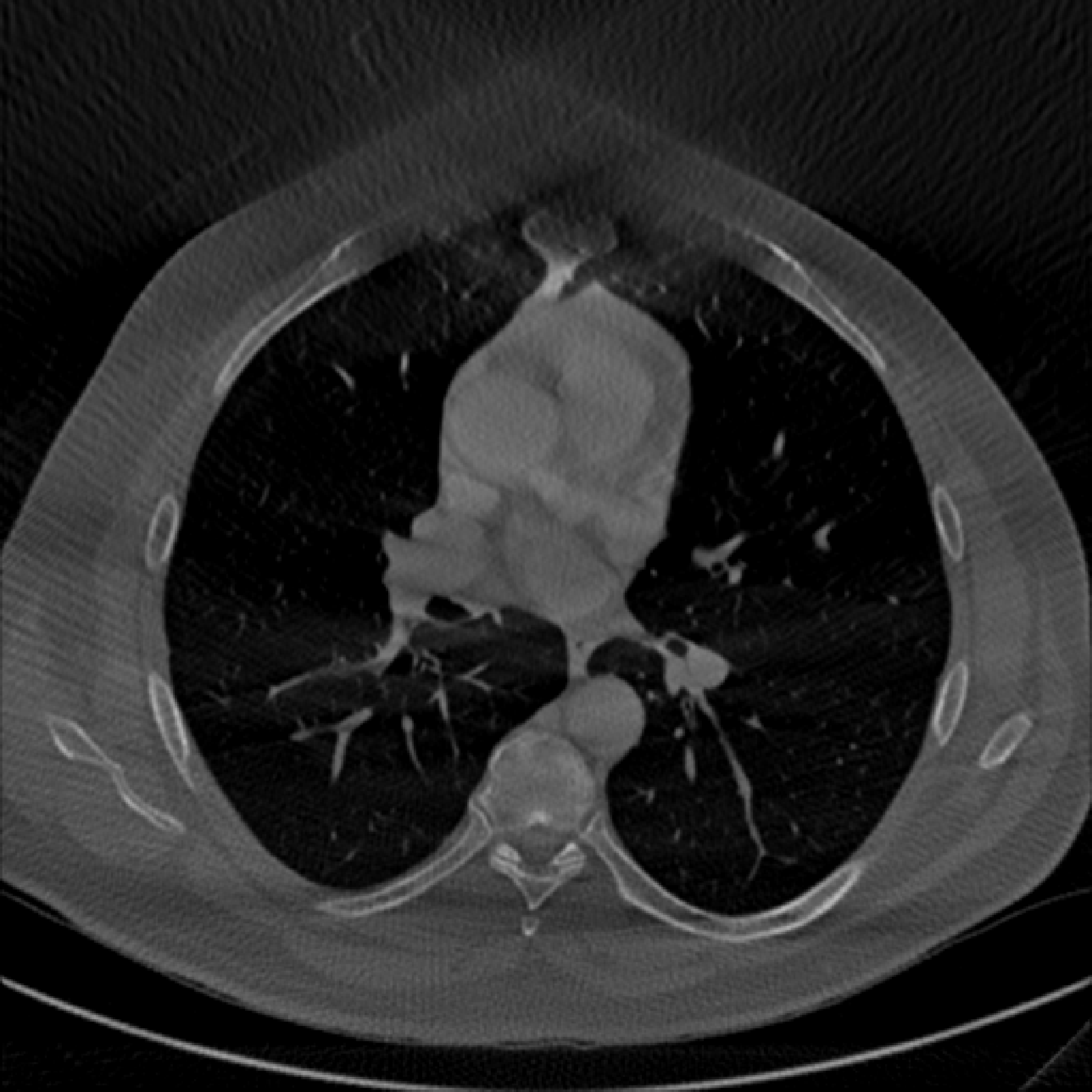}&
\subfigimg[width=\linewidth,pos=ul,font=\color{blue!90!red}]{(c)}{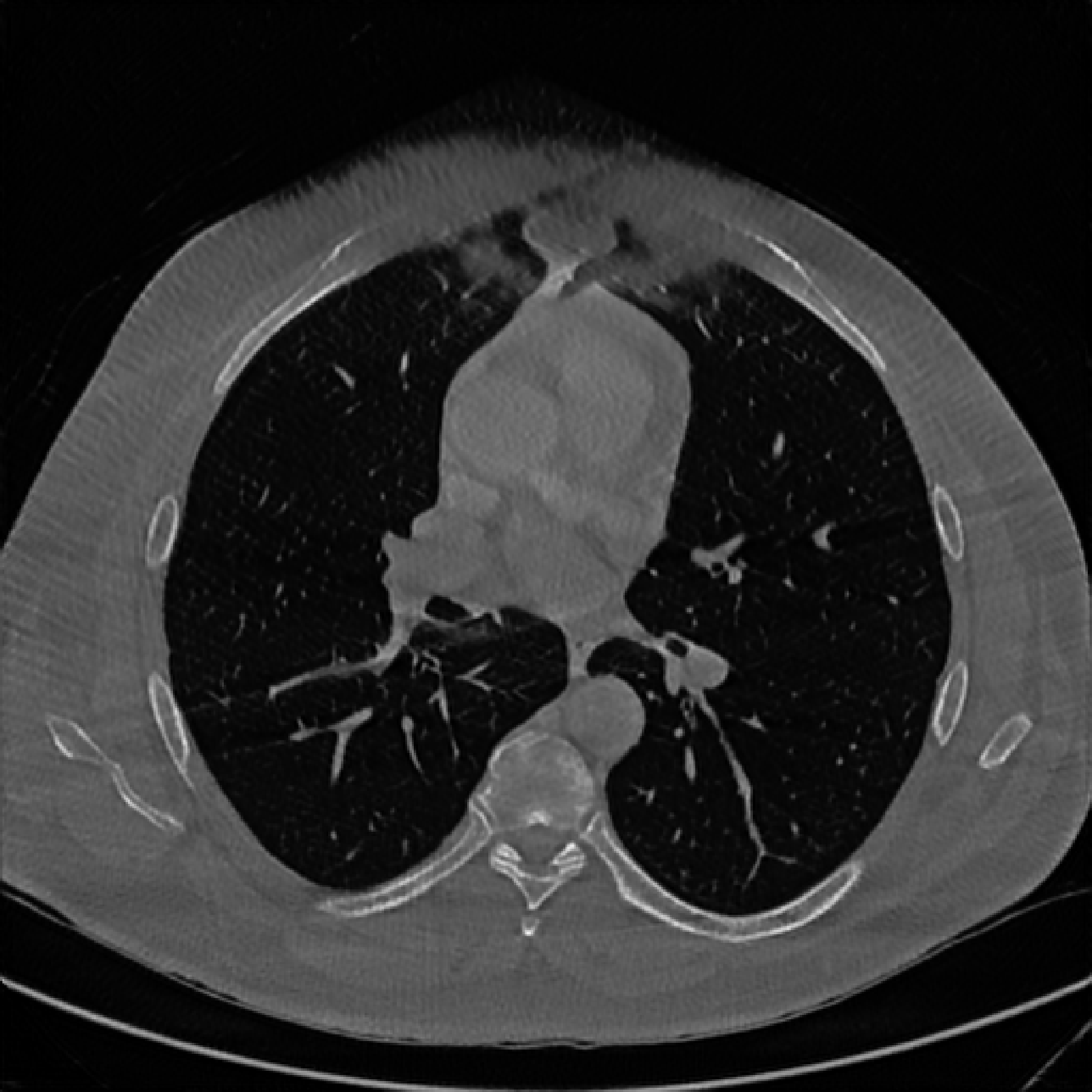}\\
\subfigimg[width=\linewidth,pos=ul,font=\color{blue!90!red}]{(d)}{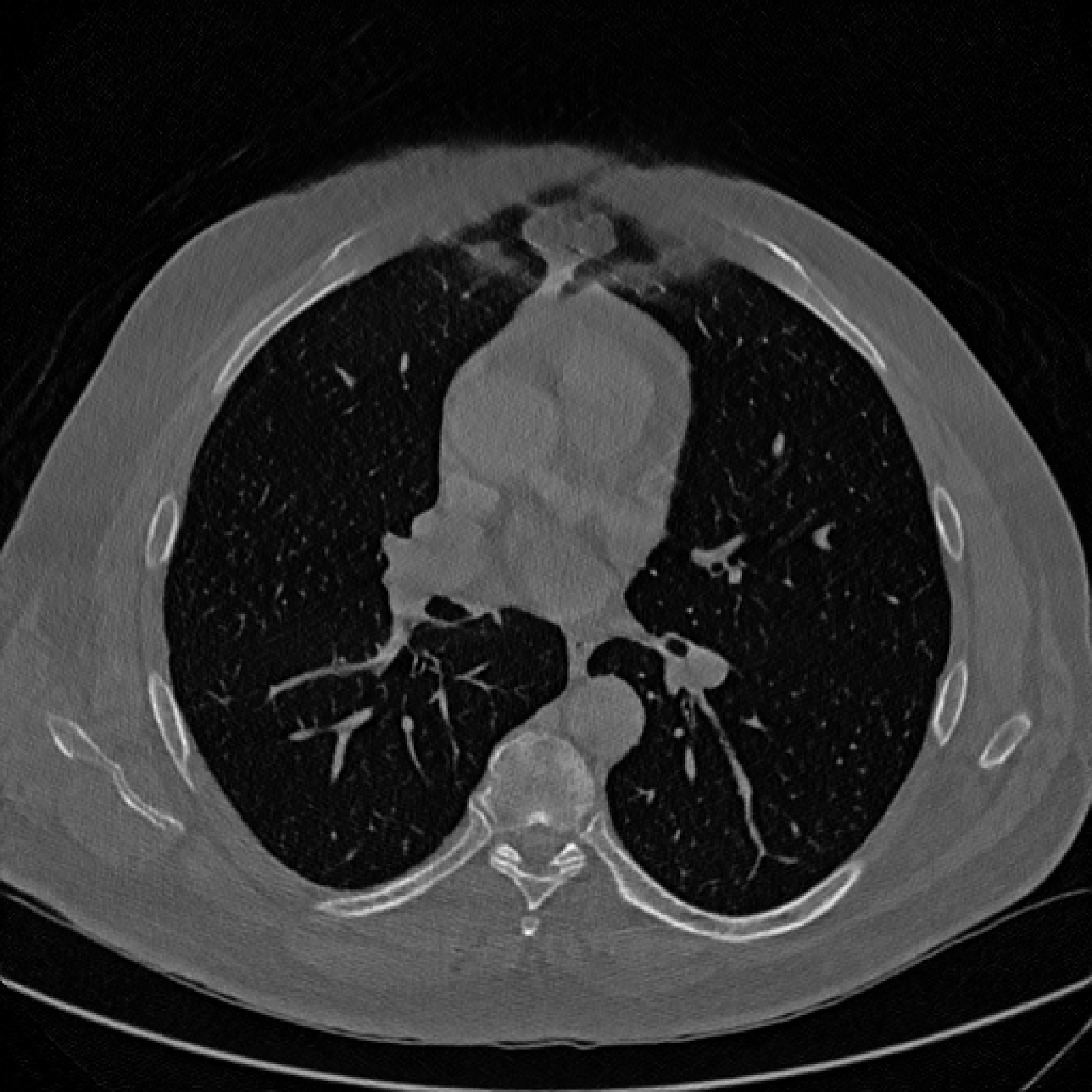}&
\subfigimg[width=\linewidth,pos=ul,font=\color{blue!90!red}]{(e)}{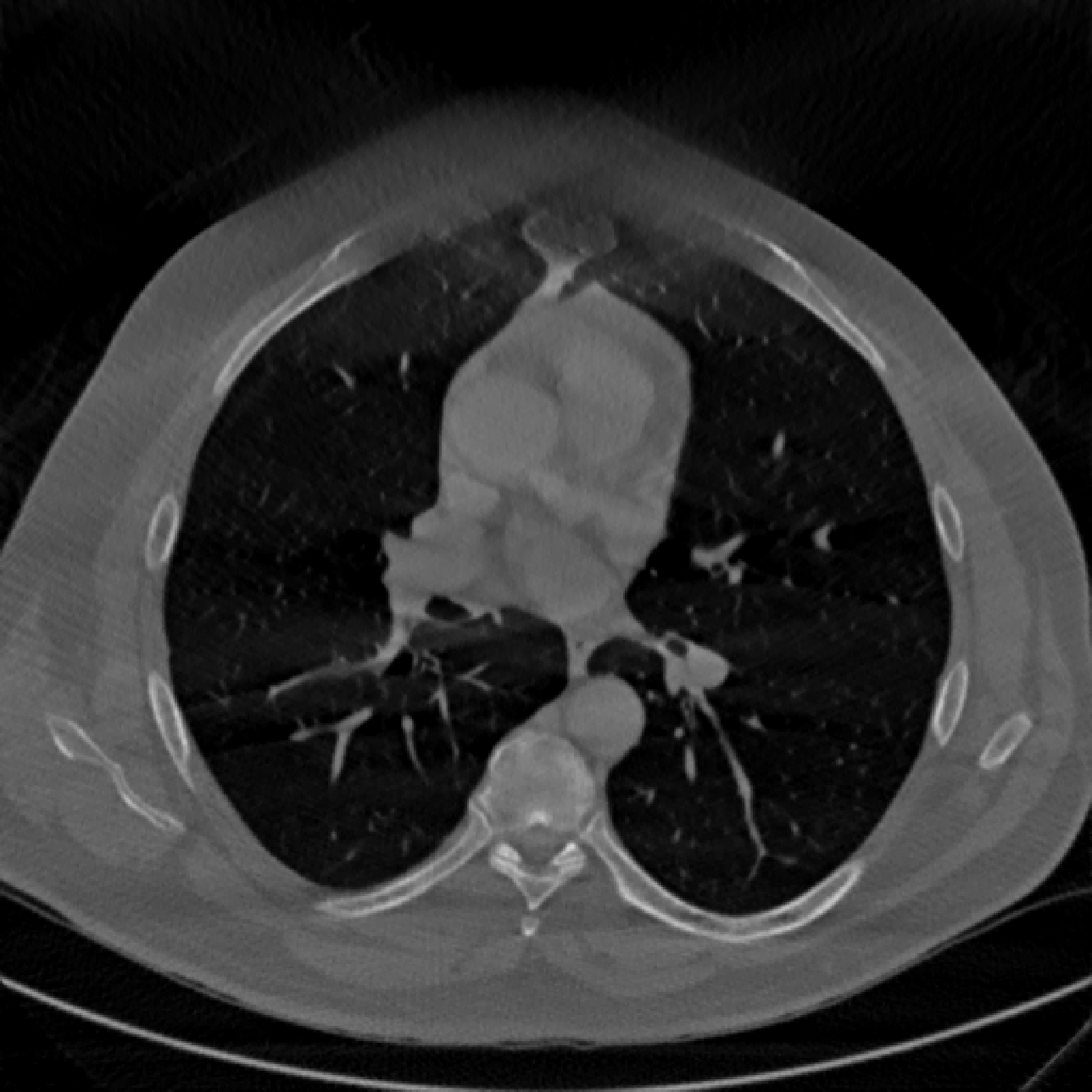}&
\subfigimg[width=\linewidth,pos=ul,font=\color{blue!90!red}]{(f)}{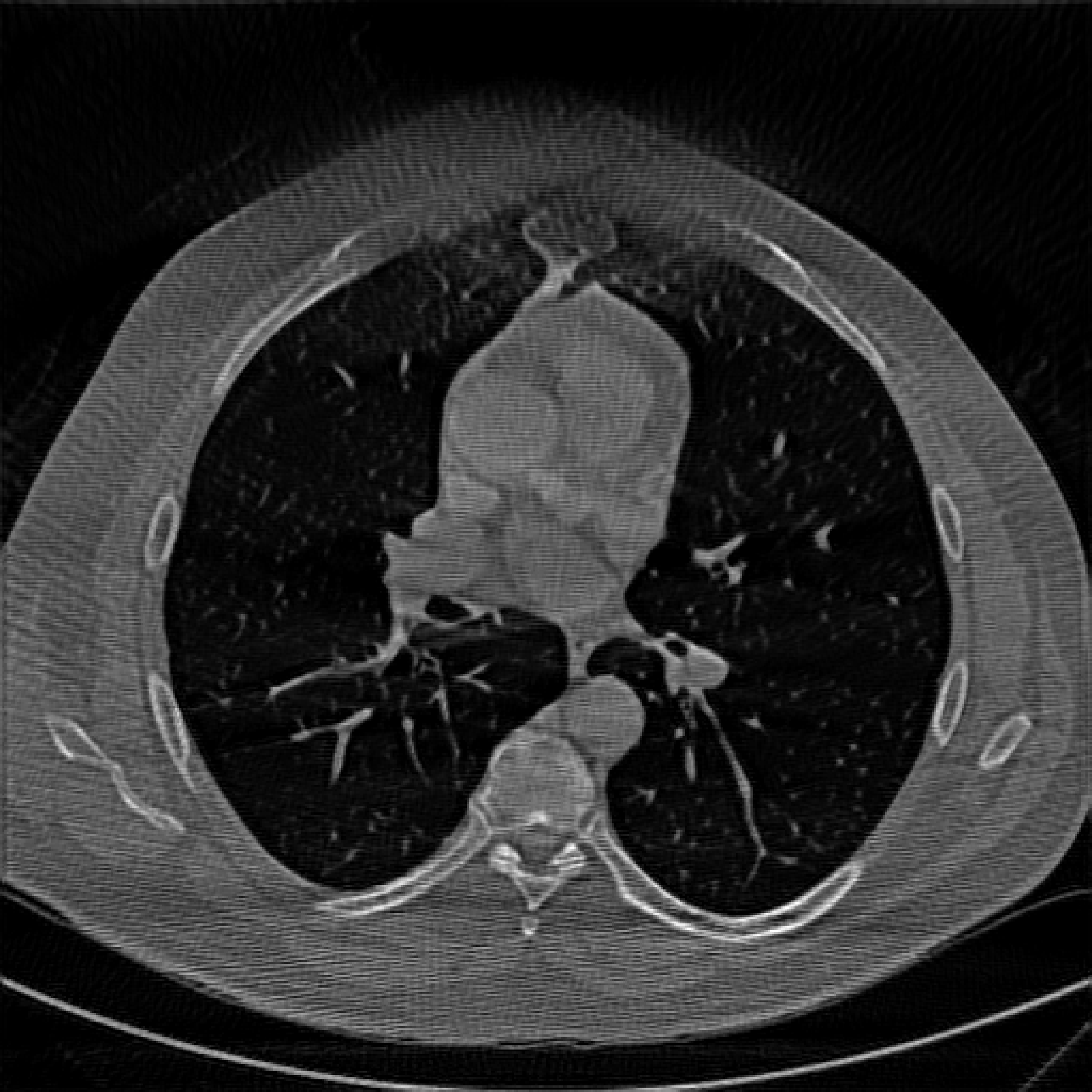}
\\
\subfigimg[width=\linewidth,pos=ul,font=\color{blue!90!red}]{(g)}{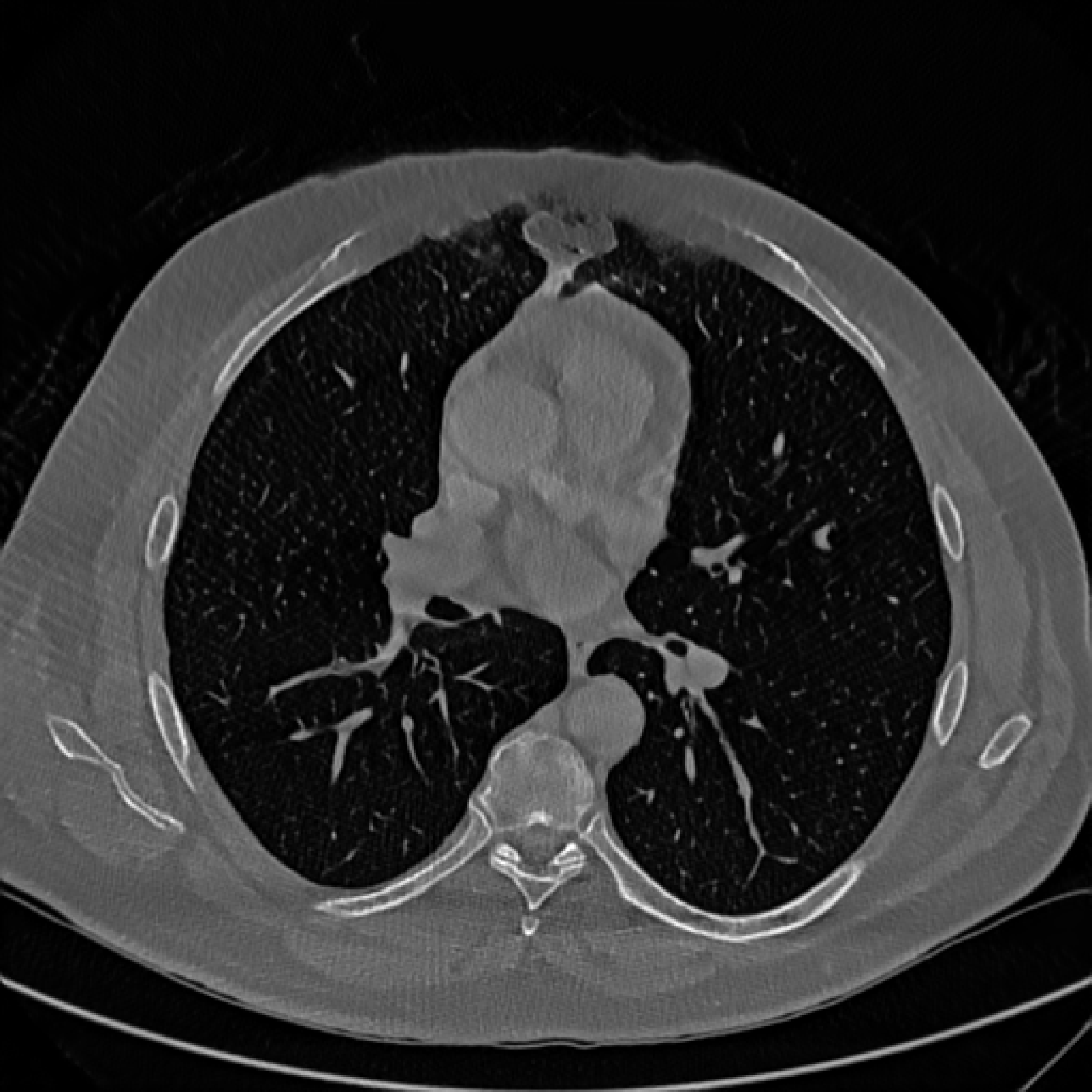}
&
\subfigimg[width=\linewidth,pos=ul,font=\color{blue!90!red}]{(h)}{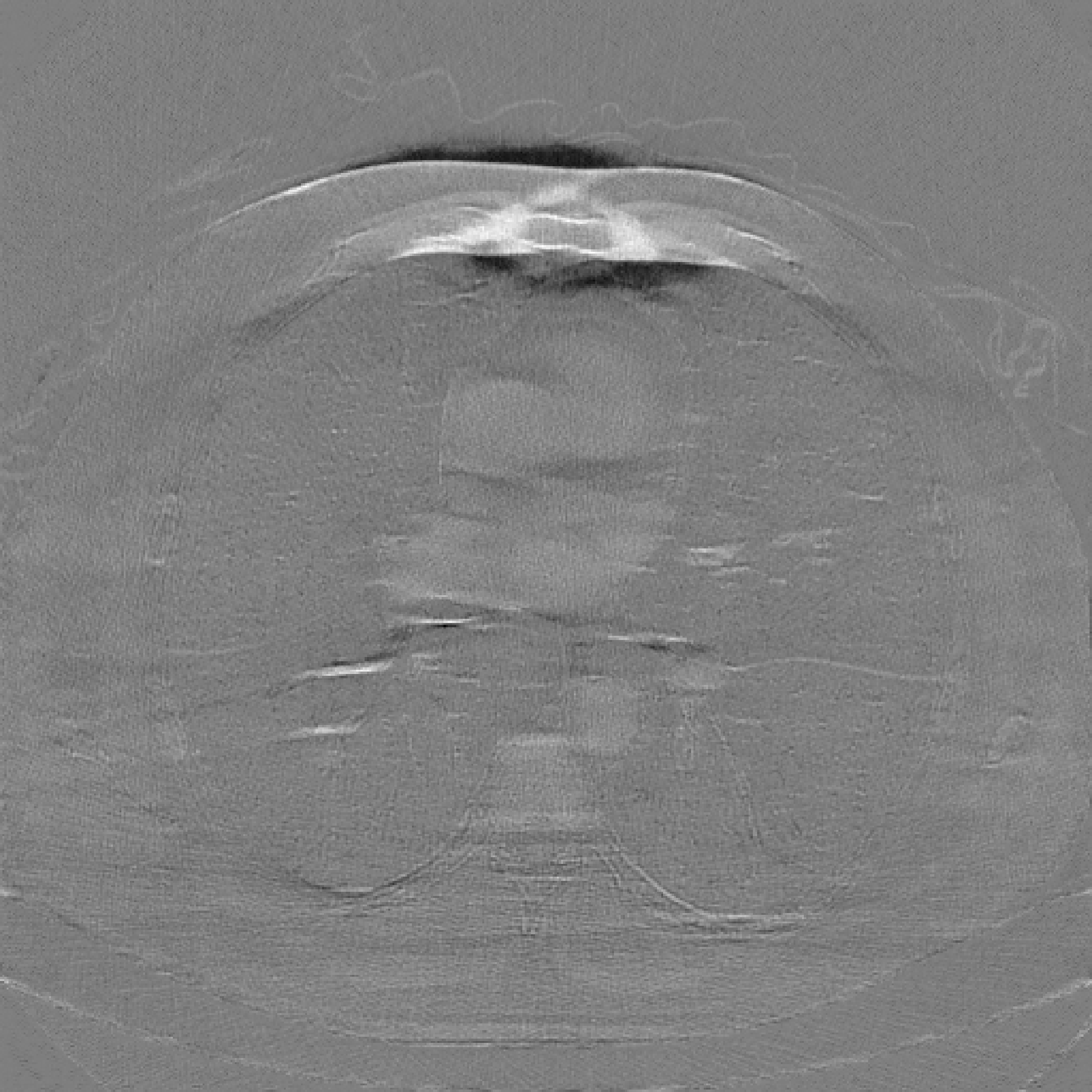} &
\subfigimg[width=\linewidth,pos=ul,font=\color{blue!90!red}]{(i)}{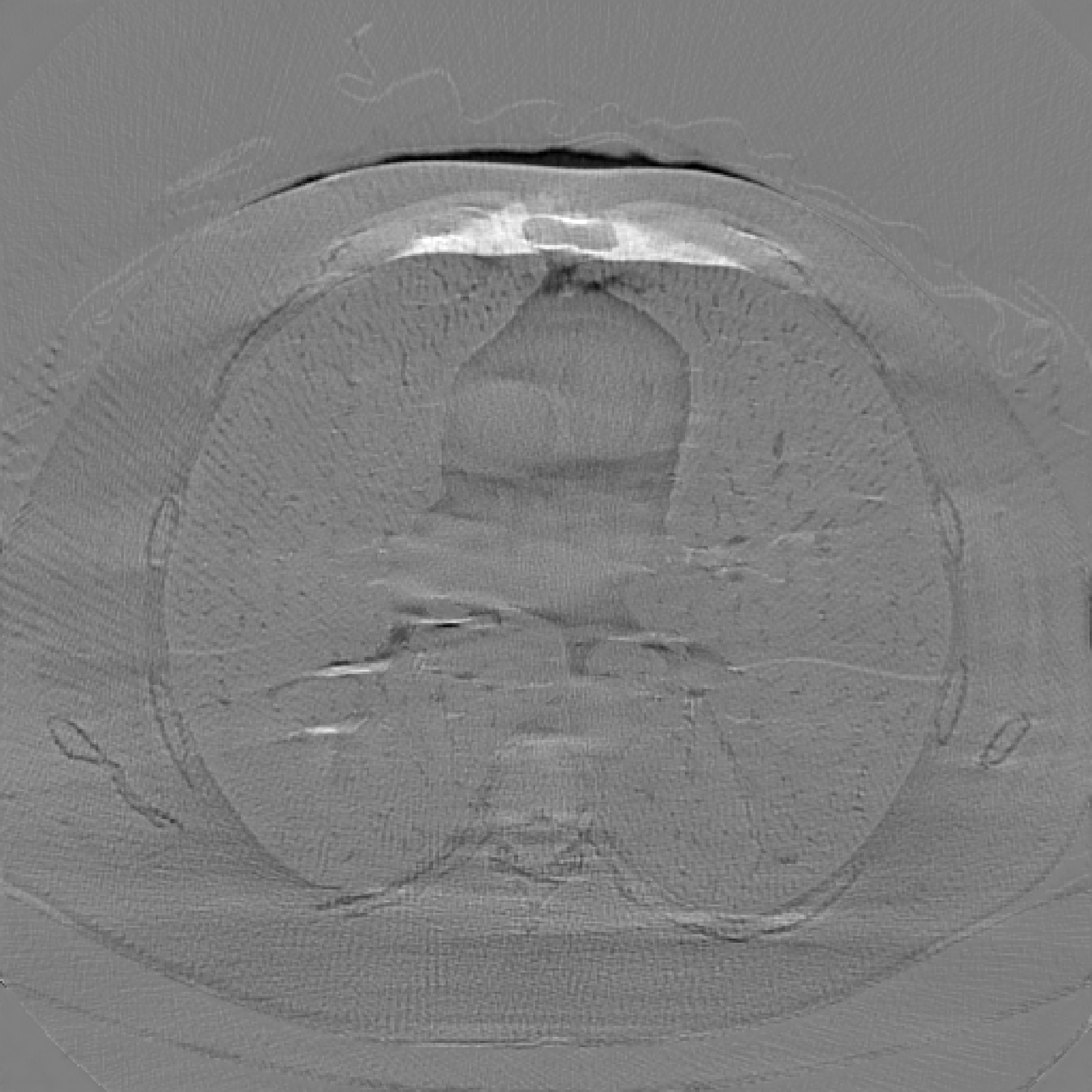}
\end{tabular}
\caption{Limited angle CT image reconstruction. (a)Ground truth; (b)FBP; (c)PD-Net, $\ell_{2}$; (d)PD-Net, SS2; (e)JSR model; (f) JSR-Net, $\ell_{2}$; (g)JSR-Net, SS2; (h)Error map of PD-Net; SS2, (i)Error map of JSR-Net, SS2.}
\label{fig:LACT}
\end{figure}

We evaluate the performance of JSR-Net for limited angle CT image reconstruction. 
Fig. \ref{fig:LACT} shows one of the example from test set with the ground truth and the reconstructed images from FBP, PD-Net \cite{adler2018learned}, JSR model(Algorithm \ref{alg:JSR-model}) and JSR-Net(Algorithm \ref{alg:JSRNet}).

The gray scale window is set the same as before except that the error map window is set to [-0.4,0.4].
Fig. \ref{fig:LACT} shows that both the PD-Net and JSR-Net outperform FBP and the JSR model, especially at the top region of the image. The use of the SS2 hybrid loss can improve the results from PD-Net over the $\ell_2$ loss but only mildly. This is because the PD-Net does not have a joint spatial Radon recovery mechanism.
In contrast, the JSR-Net can significantly benefit from the SS2 hybrid loss and it generates the best results overall compared models.
Streak artifacts are observable in Fig. \ref{fig:LACT} (c) and (f).

\begin{table*}[t]
\small
  \caption{ Incomplete data CT image reconstruction. }  \label{table:SACT_LACT}
\centering
  \begin{tabular}{|c|c|c|c|c|c|}
  \hline
\multirow{2}{*}{Tasks} & \multirow{2}{*}{Models} & \multicolumn{4}{c|}{Qual. Meas.}\\
\cline{3-6}
& & \multicolumn{1}{c|}{SSIM } & \multicolumn{1}{c|}{PSNR } &\multicolumn{1}{c|}{NRMSE} & \multicolumn{1}{c|}{MSE}\\
  \cline{1-6}
  \multirow{6}{*}{Sparse view CT}
 & FBP  & 0.6173 & 17.25 & 1.078 & 0.0189 \\
  \cline{2-6}
 & PD-Net, $\ell_{2}$  & 0.8709 & 28.54 & 0.1453 & 0.0014 \\
   \cline{2-6}
 & PD-Net, SS2 & 0.8844 & 30.68 & 0.1134 & 0.0009 \\
   \cline{2-6}
 & JSR model & 0.8088 & 26.64 & 0.1866 & 0.0022 \\
   \cline{2-6}
 & JSR-Net,$\ell_{2}$ & 0.8271 & 27.68 & 0.1604 & 0.0017 \\
    \cline{2-6}
 & JSR-Net,SS2 & \textbf{0.9081} & \textbf{31.59} & \textbf{0.1022} & \textbf{0.0007} \\
\hline
  \multirow{5}{*}{Limited angle CT}
 & FBP & 0.4826 & 15.91 & 1.5143 & 0.0257 \\
  \cline{2-6}
 & PD-Net, $\ell_{2}$ & 0.8778 & 26.43 & 0.1852 & 0.0023 \\
   \cline{2-6}
 & PD-Net, SS2 & 0.88 & \textbf{27.44} & \textbf{0.1648} & \textbf{0.0018} \\
   \cline{2-6}
 & JSR model & 0.8317 & 25.38 & 0.2174 & 0.0029 \\
    \cline{2-6}
 & JSR-Net,$\ell_{2}$ & 0.7337 & 23.72 & 0.253 & 0.0042 \\
    \cline{2-6}
 & JSR-Net, SS2 & \textbf{0.9076} & 27.31 & 0.1674 & 0.0019 \\
\hline
\end{tabular}
\end{table*}

\subsection{Quantitative comparison}
To quantitatively compare the performance of FBP, PD-Net, JSR model and JSR-Net, Table \ref{table:SACT_LACT} reports four different quality metrics of the reconstruction results: i.e., SSIM, PSNR, MSE, and NMSE. We observe that JSR-Net trained with SS2 loss outperforms PD-Net for sparse view CT and has comparable results with PD-Net for limited angle CT.

\section{Conclusion and Future work}
In this work, we propose a new CNN, named JSR-Net, by unfolding the joint spatial-Radon domain image reconstruction algorithm for incomplete data CT imaging. A hybrid loss, containing MSE, SSIM and semantic segmentation, is designed to train the proposed JSR-Net. Numerical results show improved performance of JSR-Net with hybrid loss than state-of-the-art approaches.

In the future, we will try to design more reliable loss function that is more effective in preserving details in the reconstructed image. Furthermore, the network architecture design and applications such as interior/exterior CT are also worth to explore with the proposed JSR-Net.

\end{document}